\documentclass[12pt]{iopart}
\begin{document}
\title[Hyperspherical partial wave calculation for double photoionization of helium]
{Hyperspherical partial wave calculation for double
photoionization of the helium atom at 20 eV excess energy}

\author{J. N. Das\dag,\ K. Chakrabarti\ddag\ and S.
Paul\dag}

\address{\dag\ Department of Applied Mathematics, University
College of Science, 92 Acharya Prafulla Chandra Road, Calcutta -
700 009, India}
\address{\ddag\ Department of Mathematics, Scottish Church College,
1 \& 3 Urquhart Square, Calcutta - 700 006, India}

\ead{jndas@cucc.ernet.in}

\begin{abstract}
     Hyperspherical partial wave approach has been applied here in
the study of double photoionization of the helium atom for equal
energy sharing geometry at 20 eV excess energy. Calculations have
been done both in length and velocity gauges and are found to
agree with each other, with the CCC results and with experiments
and exhibit some advantages of the corresponding three particle
wave function over other wave functions in use.
\end{abstract} \submitto{JPB}

\section{Introduction}

    There has been a significant development in the last one decade
in the theoretical study of double photoionization (DPI) of the
helium atom. This was possible because of rapid developments in
the experimental side by several groups, extending over several
countries. Double photoionization of the helium atom is one of the
most basic atomic processes. Even then, this problem merits
further detailed studies as it involves complex three body effects
and electron correlations that are not yet fully understood. Total
DPI cross section for helium has been measured by several groups
\cite{BW95,DR98,WR97,SAR98} and also calculated theoretically by
many others  \cite{PS95,KB98a,CP01,MS00}. In general there is good
agreement between the observed and the calculated cross sections.
However, there are also some discrepancies. For example
experimental results of Samson \etal and of Bizau and Wuilleumier,
although agree within 15 percent of each other, there are
systematic differences at energies above 1 Rydberg.  So also is
the case regarding the theories. Results of hyperspherical
$\mathcal{R}$-matrix with semi-classical outgoing waves
(H$\mathcal{R}$M-SOW) of Selles \etal \cite{SM02} agree favourably
with the results of Bizau and Wuilleumier \cite{BW95} while the
2SC calculation of Pont and Shakeshaft \cite{PS95} favours the
measurements of Samson \emph{et al} \cite{SAR98} (see ref.
\cite{SM02}). As regards the differential cross-sections,
understanding of the results has much improved with the
availability of results of some elaborate calculations like the
time-dependent close coupling (TDCC) calculation of Colgan
\emph{et al} \cite{CP01}, the H$\mathcal{R}$M-SOW calculation of
Malegat \etal and Selles \emph{et al} \cite{MS00,SM02} in addition
to the earlier results (\cite{BD98,KB98b,KB98c,KB00}) and the
recent results (\cite{CV00,BP01,DC01} of CCC theory, the most
extensively applied theory to the problem to date. For the status
of the CCC and some other methods the review article by Briggs and
Schimdt \cite{BS00} may be seen. Even then there remains the task
of improving the theoretical estimates, particularly at low
energies and for unequal-energy-sharing geometries (for which the
results appear comparatively less satisfactory) as well as for
better understanding of the physics of the DPI problem of helium

    In the time-independent frame work the solution of the
problem depends basically on the accurate computation of the
T-matrix element given by
\begin{equation}
T_{fi} = \langle \Psi_f^{(-)}|V|\Phi_i \rangle,
\end{equation}
where $\Phi_{i}(\vec r_1, \vec r_2)$ is the helium ground state
wave function, V is the interaction term given by
\begin{equation}
V = \vec{\epsilon} \cdot \vec{D}.
\end{equation}
$\vec D$ is the dipole operator and is given  $\vec D = \vec
\nabla_1 + \vec \nabla_2$ (velocity form) or \linebreak $\omega_i
(\vec{r}_1 + \vec{r}_2 )$ (length form) and
$\Psi_f^{(-)}(\vec{r}_1, \vec {r}_2)$ is the final channel
continuum wave function with incoming wave boundary condition for
the two outgoing electrons and $\omega_i$ is the incident photon
energy. Here $\vec\epsilon$ is the photon polarization direction
and $\vec{r}_1$, $\vec{r}_2$ are the co-ordinates of the two
outgoing electrons, the nucleus being at the origin.

    For accurate cross section results one needs accurate wave
functions $\Phi_i$ and $\Psi_f^{(-)}$. Accurate bound state helium
wave functions are easily available. There exists a number of such
wave functions for the ground state (and low-lying excited states)
in analytic form of different accuracies, such as simple Hartree
Fock type wave function used by Maulbetsch and Briggs \cite{MB93}
or a Hylleraas type wave function given by Chandrashekhar and
Herzberg \cite{CH55} or by Hart and Herzberg \cite{HH57}. If
needed, one may also readily generate an arbitrarily accurate
bound state wave function along the line developed by Pekeris
\cite{PEK58}. But for $\Psi_f^{(-)}$ there are no such simple
accurate wavefunctions in analytical form. Most wave functions
used in the literature are either incorrect at finite distances or
in the asymptotic domain, as 3C \cite{BBK89} or 2SC \cite{PS95b}
wave functions. There are now many calculations of varied
accuracies depending mainly on the use of different final channel
wave functions.

    Without caring for the explicit form of the wave functions,
Huetz and co-workers \cite{HS91} established dependence of cross
sections on various angular variables of the outgoing electrons
and on energy. For the equal energy sharing geometry case it turns
out to be rather simple in form. On least squares fitting this
gives good representation of the triple differential cross
section(TDCS) results (some authors prefer the name five fold
differential cross sections (FDCS)which is more appropriate and
relevant in view of some recent experimental results \cite{AM01}.
However, we will continue to call it TDCS).

    For the study of TDCS close to threshold there are the
Wannier calculations by Faegin \cite{FJ95, FJ96}. These give good
representation of the shape of TDCS results at 6 eV excess energy
but miserably fail at higher energies.

    There are also a number of detailed calculations by Maulbetsch
and Briggs \cite{MB93, MBS95} which used for the final state wave
function, the 3C wave function of BBK theory \cite{BBK89} and
produced moderately accurate cross section results. It is well
known that the 3C wave function is correct in the asymptotic
domain (when all of $r_1$, $r_2$ and $r_{12}$ are large) but it is
not accurate enough at finite distances. Similar calculations are
reported by Pont and Shakeshaft \cite{PS95, PS95b, PS96}. They
used screened coulomb (2SC) wave functions (for the outgoing
electrons) which is supposed to be a better wave function (but not
asymptotically exact). The results are much better.

    Later Lucey \textit{et al} \cite{LR98} tried various initial state and
final state wave functions, including the 3C wave function (none
accurate enough), tested gauge dependence and  found much
discrepancies in the results.

    In this work,  we consider yet another high level computational
scheme which is capable of yielding reliable cross sections. This
is the Hyperspherical partial wave approach of one of the authors
(Das \cite{JND94,JND98,JND01,JND02}, Das \textit{et al}
\cite{DPC03}) which is very successful in representing the
three-particle continuum wave function in the final channel very
accurately.

Here we use  a 20-term correlated Hylleraas type wave function of
the form

\begin{equation}
\Phi_i(r_1, r_2) = \aleph \,e^{-\frac{k}{2}(r_1+r_2)}
\sum_{n_1,n_2,n_3} C_{n_1 n_2 n_3}{(r_1+r_2)}^{n_1}
{(r_1-r_2)}^{n_2}\,r_{12}^{n_3}
\end{equation}

\noindent given by Hart and Herzberg \cite{HH57} for the helium
ground state and use the  final channel wave function calculated
in the hyperspherical partial wave approach as indicated below.

    It may be noted here that our present approach and the
H$\mathcal{R}$M-SOW approach are similar in some respects,
although ours is a fully quantal approach whereas the
H$\mathcal{R}$M-SOW approach is partly semi classical. From the
recent calculations of Selles \emph{et al} on H$\mathcal{R}$M-SOW
approach, it is clear that for unequal energy sharing kinematics
one needs to consider the asymptotic range parameter $R_{\infty}$
(considered below) to have values of several thousands a.u. for
converged results, whereas for equal energy sharing cases only few
hundreds suffice. So we restrict our present study to equal
energy sharing cases only, which is much easier.\\

\section{Hyperspherical Partial Wave Approach}

    In this section we outline the most salient features
of this method. For the final state $\Psi_f^{(-)}$, which needs
more accurate treatment, we use hyperspherical co-coordinates
$R=\sqrt{{r_1}^2 + {r_2}^2}$, $\alpha = atan(r_2/r_1), \;
\hat{r_1}=(\theta_{1}, \phi_{1}), \; \hat{r_2}=(\theta_{2},
\phi_{2})$ and $\omega=(\alpha, \hat{r_1}, \hat{r_2})$ and put
$P=\sqrt{{p_1}^2 + {p_2}^2}$, $\alpha_0 = atan(p_2/p_1), \;
\hat{p_1}=(\theta_{p_1}, \phi_{p_1}), \; \hat{p_2}= (\theta_{p_2},
\phi_{p_2})$ and $\omega_0=(\alpha_0, \hat{p_1}, \hat{p_2})$,
$\vec{p_1}, \; \vec{p_2}$ being momenta of the two outgoing
electrons of energies $E_1$ and $E_2$ and coordinates $\vec{r_1},
\; \vec{r_2}$ . We expand $\Psi_f^{(-)}$ in hyperspherical
harmonics (Das \cite{JND98}, Lin \cite{LIN74}) which are functions
of the above five angular variables and depend on the variables
$\ell_1,\, \ell_2, n, L, M$ (collectively called $\lambda$) which
are respectively the angular momenta of the two electrons, the
order of the Jacobi polynomial and the total angular momentum and
its projection, in addition to the dependence on S, the total
spin. It may be noted that $L, S, \pi$ (the parity) are conserved
here.

    Thus we decompose the symmetrized wave function $\Psi_{fs}^{(-)}$ as

\begin{equation}
\Psi_{fs}^{(-)}(R, \omega)=\sqrt{\frac{2}{\pi}} \sum_{\lambda}
\frac{F_{\lambda}^s(\rho)}{{\rho}^{\frac{5}{2}}}\;
\phi_{\lambda}^s(\omega),
\end{equation}
following the expansion of the symmetrized plane wave [20]
\begin{eqnarray}
 [exp(i\vec{p_1}\cdot\vec{r_1}+i\vec{p_2}\cdot\vec{r_2}) & + & (-1)^s
exp(i\vec{p_2} \cdot\vec{r_1}+i\vec{p_1} \cdot
\vec{r_2})]/(2\pi)^3\nonumber \\
& & {} =  \sqrt{\frac{2}{\pi}} \sum_{\lambda}i^{\lambda}
\frac{j_{\lambda}^s(\rho)}{{\rho}^{\frac{3}{2}}}\;
\phi_{\lambda}^{s*}(\omega_0)\;\phi_{\lambda}^{s}(\omega).
\end{eqnarray}

\noindent
Here $\lambda = \ell_1 + \ell_2 +2\,n$ and $\rho = PR$.

The $F_{\lambda}^s$ satisfy an infinite coupled set of equations
\begin{equation}
\Big[ \frac{d^2}{d\rho^2} +1 -
\frac{\nu_{\lambda}\,(\nu_{\lambda}+1)} {\rho^2}\Big]
F_{\lambda}^s (\rho) + \sum_{\lambda'} \frac{2\;\alpha_{\lambda
\lambda'}^s}{P\rho} \ F_{\lambda'}^s(\rho) = 0
\end{equation}

\noindent
where
\begin{equation}
\alpha_{\lambda \lambda'}^s = -\langle \phi_{\lambda}^s |C|
\phi_{\lambda'}^s \rangle,
\end{equation}
\begin{equation}
C=-
\frac{1}{cos\alpha}-\frac{1}{sin\alpha}+\frac{1}{|\hat{r_1}cos\alpha
-\hat{r_2}sin\alpha|}
\end{equation}

\noindent and $\nu_{\lambda}=\lambda +\frac{3}{2}$ (note that we
use $\lambda$ with two different meanings depending on the
context). Here the radial waves with L=1, S=0 and $\pi = -1$ are
relevant for the T-matrix calculations.

    So here we fix $\mu = (L,S,\pi)$ with L=1, S=0 and $\pi= -1$,
call it $\mu_0$ and consider different $N=(\ell_1, \ell_2, n)$ and
set $F_{\lambda}^s \equiv f_N^{\mu_0}$ in equation (6). Further we
omit $\mu_0$ from $f_N^{\mu_0}$ and write the relevant coupled set
of equations (6) as
\begin{equation}
\Big[ \frac{d^2}{d\rho^2} +1 - \frac{\nu_N\,(\nu_N+1)}
{\rho^2}\Big]f_N^s + \sum_{N'} \frac{2\; \alpha_{NN'}^s} {P\rho}
\, f_{N'}^s = 0.
\end{equation}
For our numerical computations we truncate the set to some maximum
value $N_{mx}$ of N. These $N_{mx}$ equations in $N_{mx}$
variables are needed to be solved from origin to infinity.
Actually we need construction of $N_{mx}$ independent solutions
which vanish at the origin. Now for convenience we divide the
whole solution domain $(0, \infty)$ into three subdomains $(0,
\Delta), (\Delta, R_{\infty})$ and $(R_{\infty}, \infty)$, where
$\Delta$ has the value of a few atomic units and $R_{\infty}$ is a
point in the asymptotic domain. Best choices for these may be made
by simple variations. Results do not depend significantly on
these. But for converged results in some situations, values of
$R_{\infty}$, as in H$\mathcal{R}$M-SOW calculation \cite{SM02},
are to be thousands of atomic units. Next we proceed for solutions
over subdomains. For $(R_{\infty}, \infty)$ we have simple
analytic solutions \cite{JND98}:
\begin{equation}
f_{snN}^s(\rho) = \sum_{\ell}\; \frac{a_{kN}^{(\ell)}sin\;
\theta_k}{\rho^\ell} +\frac{b_{kN}^{(\ell)}cos\;
\theta_k}{\rho^\ell}
\end{equation}
\begin{equation}
f_{snN}^s(\rho) = \sum_{\ell}\;
\frac{c_{kN}^{(\ell)}sin\; \theta_k}{\rho^\ell}
+\frac{d_{kN}^{(\ell)}cos\; \theta_k}{\rho^\ell}
\end{equation}

\noindent where $f^{(k)}_{snN}$ and $f^{(k)}_{csN}$ are the N-th
element of the k-th solution vectors. Obviously these give
2$N_{mx}$ independent solution vectors. The coefficients in these
expressions are determined through recurrence relations (see Das
\cite{JND98}) in terms of $a^{(0)}_{kN} = a_{kN}$ and
$b^{(0)}_{kN} = 0$, $c^{(0)}_{kN} = 0$, $d^{(0)}_{kN} = a_{kN}$,
$a_{kN}$ being the N-th element of the k-th eigen vector of the
charge matrix $A=(\alpha_{NN'})$. Here we have $\theta_k = \rho +
\alpha_k ln\;2\rho$, $\alpha_k$ being the k-th eigen value of A.

    Solution over $(\Delta, R_{\infty})$ is also very simple.
Because of the simple structure of equations (9) a Taylors
expansion method works nicely. Earlier for the (e, 2e) problem,
Das also adopted this approach \cite{JND01,JND02}. But the main
difficulty lies in the construction of the solution vectors over
$(0, \Delta)$. In those calculations on (e, 2e) problem Das used
an approach as in R-matrix calculations \cite{BR75}. But very
often, this invites pseudo resonance type behaviour causing
undesirable oscillations in the cross sections. Recently \cite
{DPC03} for \linebreak (e, 2e) problem, we applied the finite
difference method (a five-point scheme) for solutions in the
interval $(0, \Delta$) and thereby get rid of undesirable
oscillations. So we adopted here the same approach but with a
seven-point scheme in place of the five-point scheme.

    Thus for the solution in the interval $(0, \Delta)$ we recast
equations (9) in terms of R instead of $\rho$, as
\begin{equation}
\Big[ \frac{d^2}{dR^2} + P^2 - \frac{\nu_N\,(\nu_N+1)}
{R^2}\Big]f_N^s + \sum_{N'=1}^{N_{mx}} \frac{2\;\alpha_{NN'}^s}
{R} \,f_{N'}^s = 0,
\end{equation}

\noindent and solve these equations as a two point boundary value
problem by difference equation method. At $R = 0$, the solution
vectors are set to zero while at $R=\Delta$ we assign to the k-th
solution vector the k-th column of the unit matrix. The matrix for
the corresponding difference equation is a sparse matrix and for
its solution special methods are available. Here we use
biconjugate gradient method \cite{FL75}. We find that this method
readily works and gives converged solutions.

    Now for the difference equations we divide the interval $[0, \Delta]$
into \textit{m} subintervals of length h with mesh points $$0=R_0
\; < \; R_1 < \; R_2 \; < \; \cdots < \;R_k < \cdots < \; R_{m-1}
\; < R_m=\Delta$$ with $R_k = R_0 + kh$ and use the following
seven-point difference formula:\pagebreak

\begin{eqnarray}
f_N^{''}(R_k)&=&\frac{1}{\,h^2}\Big[\ \frac{1}{90}f_N(R_{k-3})-
\frac{3}{20}16f_N(R_{k-2}) \nonumber\\
&+& \frac{3}{2}f_N(R_{k-1})-\frac{49}{18} f_N(R_{k})
+\frac{3}{2}f_N(R_{k+1}) - \frac{3}{20}16f_N(R_{k+2}) \nonumber\\
&+&\frac{1}{90}f_N(R_{k+3})\Big]
+\{\frac{69}{25200}\,h^6f_N^{(viii)}(\xi_1)\}
\end{eqnarray}
for $k = 3, 4, \cdots , m-4, m-3$, and for $k = 1, 2$ and $m-2,
m-1$ the formulae
\begin{eqnarray}
f_N^{''}(R_1) &=&\frac{1}{h^2}\Big[ \frac{3}{8}f_N(R_0)+6f_N(R_1)-
\frac{11}{2}\,h^2f_N^{''}(R_2) \nonumber\\
&-& \frac{51}{4}f_N(R_3)-\,h^2f_N^{''}(R_3)+6f_N(R_4
+\frac{3}{8}f_N(R_4)\Big]\nonumber\\
&+& \{-\frac{23}{10080}\,h^6f^{(viii)}(\xi_2)\}.
\end{eqnarray}
\begin{eqnarray}
f_N^{''}(R_2) &=&\frac{1}{h^2}\Big[ \frac{3}{8}f_N(R_1)+6f_N(R_2)-
\frac{11}{2}\,h^2f_N^{''}(R_3) -
\frac{51}{4}f_N(R_3) \nonumber\\
&-&\,h^2f_N^{''}(R_4)+6f_N(R_4)+\frac{3}{8}f_N(R_5)\Big]\nonumber\\
&+& \{-\frac{23}{10080}\,h^6f^{(viii)}(\xi_3)\}.
\end{eqnarray}
\begin{eqnarray}
f_N^{''}(R_{m-2}) &=&\frac{1}{h^2}\Big[
\frac{3}{8}f_N(R_{m-5})+6f_N(R_{m-4})-\,h^2f_N^{''}(R_{m-4})
-\frac{51}{4}f_N(R_{m-3})\nonumber\\
&&-\frac{11}{2}\,h^2f_N^{''}(R_{m-3})
+6f_N(R_{m-2})+\frac{3}{8}f_N(R_{m-1})\Big]\nonumber\\
&&+ \{-\frac{23}{10080}\,h^6f^{(viii)}(\xi_4)\}.
\end{eqnarray}
    and
\begin{eqnarray}
f_N^{''}(R_{m-1}) &=&\frac{1}{h^2}\Big[
\frac{3}{8}f_N(R_{m-4})+6f_N(R_{m-3})-\,h^2f_N^{''}(R_{m-3})-
\frac{51}{4}f_N(R_{m-2})  \nonumber\\
&&-\frac{11}{2}\,h^2f_N^{''}(R_{m-2})+6f_N(R_{m-1})+\frac{3}{8}f_N(R_{m})\
\Big]\nonumber\\
&&+ \{-\frac{23}{10080}\,h^6f^{(viii)}(\xi_5)\}.
\end{eqnarray}

The quantities on the right hand sides within curly brackets
represent the error terms. The corresponding difference equations
are obtained by substituting these expressions the values of
second order derivatives from the differential equation (12). For
continuing these solutions in the domain $(\Delta, \; R_\infty)$
we need first order derivatives ${f'}_N(R)$ at $\Delta$. These are
computed from the difference formula

\begin{eqnarray}
f_N^{'}(R_m) & = & \frac{1}{84h}[-f_N(R_{m-4} +
24f_N(R_{m-2})-128f_N(R_{m-1})+105f_N(R_m)] \nonumber \\
& + & \frac{2h}{7}f_N^{''}(R_n) +
\{-\frac{4h^4}{105}f_N^{(v)}(\xi)\}
\end{eqnarray}

    Here too, the quantity within curly brackets represents the
error term. The solutions thus obtained in $(0, \Delta)$ are then
continued over $(\Delta, R_{\infty})$ by Taylor's expansion
method, as stated earlier, with stabilization after suitable steps
\cite{CT75}. The $N_{mx}$ independent solution vectors so
obtained, are put together to get the solution matrix $f_0$. The
solution matrices $f_{sn}$ and $f_{cs}$ are similarly obtained,
whose N-kth element are respectively $f^{(k)}_{snN}$ and
$f^{(k)}_{csN}$, given by (10) and (11) respectively.

    Next we introduce the K-matrix through the relation
\begin{equation}
f_0 \cdot B = f_{sn} +f_{cs} \cdot K
\end{equation}

\noindent where $B$ is an unknown constant matrix. The K-matrix is
determined from matching values and first order derivatives at
$R_\infty$, where all of $f_0, \; f_{sn}$ and $f_{cs}$ are valid.
(It may be noted here that there is a slight departure in our
definition of K-matrix from the usual practice. However, it is
symmetric as it should be).

    Finally the physical scattering state with appropriate boundary
conditions is taken as

\begin{equation}
\mathbf{f_{ph}} = f_0 \cdot \textbf{g}
\end{equation}

\noindent
and also we have
\begin{eqnarray}
\mathbf{f_{ph}} & = & (f_{sn} + f_{cs} \cdot K) \cdot
\mathbf{c}  \nonumber\\
       & = & f_{sn}\cdot \mathbf{c} + f_{cs} \cdot \mathbf{d}
\end{eqnarray}
with
\begin{equation}
\mathbf{d} = K \cdot \mathbf{c}.
\end{equation}

\noindent Thus the physical state is completely determined once
the vector $\mathbf{c}$ is determined. Now $\mathbf{c}$ is
determined from the consideration that $\Psi_{fs}^{(-)}$ is
asymptotically a (distorted) plane wave (representing the two
outgoing electrons) plus incoming waves only. So the coefficients
of the outgoing wave $exp(i\rho)$ of both $\Psi_{fs}^{(-)}$ and
the symmetrized plane wave (equation (5)) must be the same (except
for the distorting term $exp(i\alpha_k ln\;2\rho)$). This requires

\begin{equation}
\mathbf{c} = [I + iK]^{-1}\mathbf{P}
\end{equation}

\noindent
where
\begin{equation}
\mathbf{P} = -2 e^{i\frac{\pi}{4}} \;
X^{-1}\mathbf{{\Phi}^{s*}(\omega_0)},
\end{equation}
\noindent
 and X is the matrix comprising of the columns of eigen vectors of the
charge matrix A and $\mathbf{{\Phi}^{s*}}$ is given by

\begin{equation}
\mathbf{ {\Phi}^{s*}(\omega_0)}=
\left(\begin{array}{c}
\phi_1^{s*}(\omega_0)\\
\; \vdots\\ \phi_{N_{mx}}^{s*}(\omega_0)
\end{array} \right).
\end{equation}

    Finally the DPI triple differential cross section is given by
\begin{equation}
\frac{d^3\sigma}{d\Omega_1 d\Omega_2 dE_1} = \frac{2\pi^2\alpha
p_1p_2}{\omega_i} |T_{fi}|^2,
\end{equation}
after the inclusion of $\mu_0$-part of $\Psi_{fs}^{(-)}$ in
$T_{fi}$.

\section{Results}

   In our present calculation we have applied the above
hyperspherical partial wave approach both in length and velocity
gauges. Here we consider equal-energy-sharing case, since then the
computational problem becomes little simple. For this calculation
we have chosen $\Delta = 5$ a.u., $R_{\infty} = 300$ a.u., $h =
0.05$ a.u. upto $\Delta$ and 0.1 a.u. beyond $\Delta$. We have
included 90 coupled channels with $n$ upto 9 and $(l_1, l_2)$
combinations nearly as in ECS calculation \cite{BR01} for electron
- hydrogen ionization collision. We have chosen the case of
ionization at 20 eV excess energy as it has been widely considered
and for which there are interesting experimental results
\cite{BD98}. For the present calculations with 90 channels and
$R_{\infty} = 300$ a.u., our single differential cross section
(SDCS) is little above the desired value of about 0.93 Kb/eV at
$E/2$ (E being the excess photon energy). So we normalized our
TDCS by scaling with a factor 0.8 (which is also the factor we use
to scale our SDCS to get the desired value of 0.93 Kb/eV at $E/2$)
both in the length and in the velocity gauges. The TDCS results
thus obtained are presented in \linebreak figure 1. Here we
compare our results with the experimental results of Br\"auning
\textit{et al} \cite{BD98} and with the theoretical results of the
CCC calculation only \cite{BD98,ASK}, since the overall agreement
of the CCC results are known to be somewhat better compared to the
results of other calculations like TDCC, H$\mathcal{R}$M-SOW etc.
In all the cases the agreement between the velocity and length
gauge calculations is excellent everywhere. We also did
calculations in the acceleration gauge and these are
indistinguishable from those of the velocity gauge. Agreement with
the experimental results and with the CCC results are also
generally good except for some spurious peaks at $\theta_{p_1} =
0^o$. The CCC results appear a little better compared to ours.

    For unequal energy sharing our approach also works and we have
reasonably good results. But for fully converged results, we have
to consider larger values for $R_{\infty}$. So we wish to report
such results in future when our study is completed.

\section{Conclusions}

    The present calculation, reported here, has only approximately
converged. The results we have obtained, go to show that the
hyperspherical three-particle scattering state wave function, used
in the present calculation, must be  reasonably accurate from
small distances to the asymptotic region, since the results in all
the three gauges are practically identical. In contrast, the 3C or
other similar wave functions, which are not accurate at finite
distances, show strong gauge dependence \cite{LR98}. We also
mention that the present calculation is free from any genuine
difficulty and does not show any weakness worth mentioning. The
present approach may easily be applied to DPI with varied types of
polarization of the incident photons. With judicial choice of the
parameter $R_{\infty}$ and possibly with the availability of
better computational facilities, the method may be applied from
very low energy to high energy cases. At this  point if we recall
the capability of the hyperspherical partial wave approach in
representing electron-hydrogen-atom ionization collisions
\cite{JND01,JND02,DPC03} at low energies (and also consider
situations of very low energy cases, with excess energy 1 eV and
below for which we had to take $R_{\infty}$ about 4000 - 5000 a.u.
and get reliable results for ionization cross sections
\cite{DP03}), consider the present success, then we may expect the
hyperspherical partial wave theory to have a very good prospect.

\ack

    We are grateful to H. Br\"auning for providing us with the
experimental results and to Igor Bray and Anatoly Kheifets for
providing us with the CCC results in electronic form. KC
acknowledges support from the UGC in the form of a  Minor Research
Project F.PSW-035/02(ERO). SP is grateful to CSIR for providing a
research fellowship. \pagebreak


\pagebreak

\begin{center}
\Large{\underline{Figure Captions}}
\end{center}

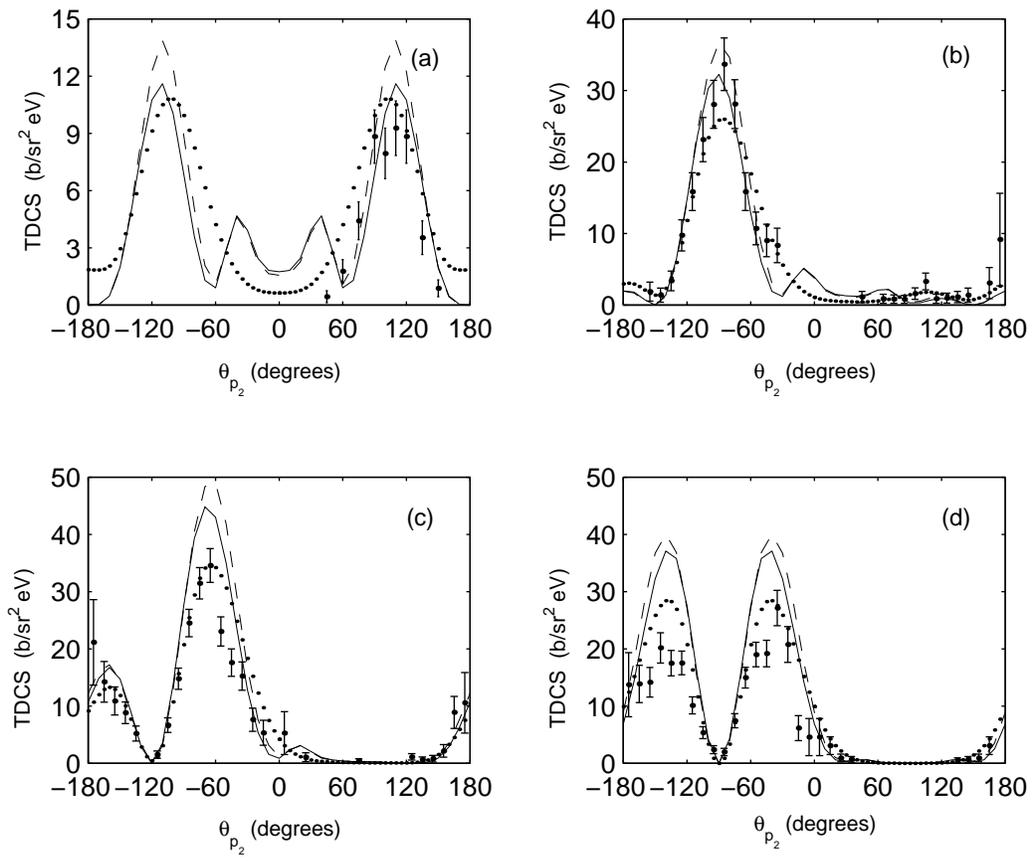
\noindent \textbf{Figure 1.} Triple differential cross sections
for photo double ionization of the helium atom for equal energy
sharing geometry for 20 eV excess energy and for a) $\theta_{p_1}
= 0^o$, b) $\theta_{p_1} = 30^o$, c) $\theta_{p_1} = 60^o$, d)
$\theta_{p_1} = 90^o$, $\theta_{p_1}$ being measured from the
photon polarization direction. Theory : continuous curve, present
calculation in velocity gauge; dashed curve present calculation in
length gauge; dotted curve, CCC calculation \cite{BD98};
Experiment : absolute measured values of Br\"auning \textit{et
al}\cite{BD98}.

\end{document}